\documentstyle[12pt]{article}

\newcommand{\newsection}{    
\setcounter{equation}{0}
\section}
\newcommand{\tr}[1]{\,{\rm tr}\,#1\,}
\def\e{{\,\rm e}\,}

\def\eop{\vspace*{\fill}\pagebreak}
\def\be{\begin{equation}}
\def\ee{\end{equation}}
\def\bea{\begin{eqnarray}}
\def\eea{\end{eqnarray}}

\newcommand{\rf}[1]{(\ref{#1})}
\newcommand{\eq}[1]{Eq.~(\ref{#1})}

\def\bb{\bar{\beta}}

\def\l{\lambda}
\def\h{\eta}
\newcommand{\ie}{{\it i.e.}\ }

\renewcommand{\d}{{{\partial}}}
\newcommand{\p}{{\prime}}
\newcommand{\ra}{\rightarrow}
\newcommand{\fr}[2]{{\textstyle {#1 \over #2}}}

\begin{document}

\begin{titlepage}
\begin{flushright}
ITEP-YM-11-92 \\ December, 1992
\end{flushright}
\vspace{1cm}

\begin{center}
{\LARGE Matrix Models of Induced Large-$N$ QCD} \end{center} \vspace{.5cm}
\begin{center}
{\large Yu.\ Makeenko}\footnote{E--mail: \ makeenko@vxitep.itep.msk.su \ / \
makeenko@desyvax.bitnet \ / \ makeenko@nbivax.nbi.dk \ }
\\ \mbox{} \\
{\it Institute of Theoretical and Experimental Physics} \\ {\it
B.Cheremuskinskaya 25, 117259 Moscow}
\end{center}

\vspace{1cm}

\begin{abstract}
I review recent works on the problem of inducing large-$N$ QCD by matrix
fields.
In the first part of the talk I describe the matrix models which induce
large-$N$ QCD and present the results of studies of their phase structure
by the standard lattice technology (in particular, by the mean field method).
The second part is devoted to the exact solution of these models
in the strong coupling region by means of the loop equations.
\end{abstract}

\vspace{1.5cm}
\noindent
Talk at the Seminar on Strings, Matrix Models and all that,
Rakhov, FSU, October 1992.

\eop
\end{titlepage}
\setcounter{page}{2}

\newsection{Introduction}

Recently there has renewed interest in the problem of inducing QCD by means
of some pre-theory. As was proposed by Kazakov and Migdal~\cite{KM92},
such a theory can be potentially solvable in the limit of large number of
colors, $N_c$, providing the inducing model is that of the
(self-interacting) matrix scalar field in the {\it adjoint\/}
representation of the gauge group $SU(N_c)$ on the lattice.
The gauge field is attached in the usual way to
make the model gauge invariant except no kinetic term for the gauge field.
The latter circumstance differs the Kazakov--Migdal model from the
standard lattice Higgs--gauge models. It was conjectured in Ref.~\cite{KM92}
that the model undergoes, with decreasing the bare mass of the scalar field,
a second order phase transition which is associated with continuum QCD
when the critical point is approached from the strong coupling region.

To solve the Kazakov--Migdal model in the strong coupling region,
Migdal~\cite{Mig92a} applied
the Riemann--Hilbert method and derived the master field
equation to determine the $N_c=\infty$ solution.
An explicit solution of this equation for the quadratic potential is
found by Gross~\cite{Gro92}. A surprising property of the master field
equation (not yet completely understood) is that it
admits~\cite{Mig92a,Mig92c} self-consistent scaling solutions
with non-trivial critical indices for the non-quadratic potentials.
Moreover, the very Riemann--Hilbert method of Ref.~\cite{Mig92a} was
developed, in fact, to find such a scaling solution.

These nice features of the Kazakov--Migdal model are due to the fact
that the scalar field is in the adjoint representation of the gauge
group. For this reason it can be diagonalized by a (local) gauge
transformation so that only ${\cal O}(N_c)$ degrees of freedom are left and
the saddle-point method is applicable as $N_c\ra\infty$.
However, a
 price for having the adjoint-representation field is an extra local $Z_N$
symmetry which leads~\cite{KSW92} at $N_c=\infty$ to {\it local
confinement\/}
(\ie the infinite string tension) rather than area law for the Wilson loops.

In the present talk I review the papers~\cite{KhM92a,Mak92,KhM92b} where
the questions of which models induce large-$N_c$ QCD with normal area and
how to solve these models in the strong coupling region were answered%
\footnote{For a review of other approaches, see recent
surveys~\cite{Kaz92,Gro92b,SW93} and references therein}.  In Sect.~\ref{Mamo}
I describe the models, both scalar and fermion ones, which induce large-$N_c$
QCD. The mechanism exploited is based on the first order phase transition which
occurs  with decreasing bare mass of the inducing field and is associated with
freezing the gauge field and the restoration of area law.  In
Sect.~\ref{loops} I discuss the exact strong coupling solution for the
quadratic potential, both in scalar and in fermion cases.  It is obtained by
solving loop equations which turn out to be a useful tool for studies of
the matrix models.

\newsection{The models of induced large-$N$ QCD \label{Mamo}}

Since the inducing matter field is in the adjoint representation of $SU(N_c)$,
Wilson loops vanish to each order of the large mass expansion at $N_c=\infty$.
This situation is associated with local confinement. However, the area law is
restored with decreasing bare mass at the point of the first order large-$N_c$
phase transition. Its existence can be rigorously proven for the
single-plaquette adjoint action and is expected for more complicated models on
the basis of the mean field method. Once the first order phase transition
occurs, the proper model will induce large-$N_c$ QCD in the continuum.

\subsection{Adjoint scalar model}

A simplest way to induce large-$N_c$ QCD is by adjoint representation scalars
at $N_f$ flavors (\ie the number of different species). The {\it adjoint scalar
model\/} ASM is defined by the partition function
\be
Z_{ASM}=\int \prod_{x,\mu} dU_{\mu}(x) \prod_x
\prod_{f=1}^{N_f} d\Phi_f(x) \e^{\sum_{f=1}^{N_f} \sum_x N_c
\tr{\left(-V[\Phi_f(x)]+
\sum_{\mu=1}^D\Phi_f(x)U_\mu(x)\Phi_f(x+\mu)U_\mu^\dagger(x)\right)}}
\label{spartition}
\ee
where the fields $\Phi_f(x)$ ($f=1,\ldots,N_f$) take values in the
adjoint representation of the gauge group $SU(N_c)$ and the link variable
$U_\mu(x)$ belongs to the gauge group. The potential $V[\Phi]$ is given by
\be
V[\Phi]=\frac m2 \Phi^2 +\ldots
\label{potential}
\ee
where $m$ is the (square of the) bare mass of the scalar field.
The original Kazakov--Migdal model~\cite{KM92} corresponds to $N_f=1$.
Notice that the action in \eq{spartition} is diagonal w.r.t.\ the flavor
indices.

The case of $N_f=1$ is a unique one when the matrix $\Phi(x)$ can be reduced
to a diagonal form at each site of the lattice by means of a gauge
transformation. Only at $N_f=1$ the Riemann-Hilbert method of
Ref.~\cite{Mig92a} is therefore applicable. The
alternative method of solving ASM with the quadratic potential at strong
coupling is based on loop equations~\cite{Mak92} and can be used at any $N_f$
while gives at $N_f=1$ the same result as the Riemann--Hilbert method.

It is worth noting that one can integrate in~\rf{spartition} over arbitrary
hermitean matrices rather than over those in the adjoint representation which
gives at finite $N_c$ a different model for the case of a non-quadratic
potential. These two modeld should coincide, however, as $N_c\ra\infty$.

\subsection{Adjoint fermion model}

An alternative to ASM is
the {\it adjoint fermion model} (AFM) which is defined by
the partition function
\be
Z_{AFM}=\int \prod_{x,\mu} dU_{\mu}(x) \prod_x \prod_{f=1}^{N_f} d\Psi_f(x)
d\bar{\Psi}_f(x) \e^{-S_{F}[\Psi,\bar{\Psi},U]}.
\label{fpartition}
\ee
Here $S_{F}[\Psi,\bar{\Psi},U]$ is the lattice fermion action
\bea
 & & S_{F}[\Psi,\bar{\Psi},U] =
\sum_{f=1}^{N_f} \sum_x N_c \tr{\Big(V_{even}({\bar{\Psi}_f(x)\Psi_f(x)})
\nonumber \\* & & -
\sum_{\mu=1}^D [\bar{\Psi}_f(x)P_\mu^-U_\mu(x)\Psi_f(x+\mu)U_\mu^\dagger(x)}
+\bar{\Psi}_f(x+\mu)P_\mu^+U_\mu^\dagger(x)\Psi_f(x)U_\mu(x)]\Big)
\label{faction}
\eea
where
\be
V_{even}({\bar{\Psi}\Psi}) = m
\bar{\Psi}\Psi \label{fpotential} +\ldots
\ee
is a fermionic analogue
of the potential~\rf{potential} and $m$ is the bare mass of the fermion
field.

In Eqs.~\rf{fpartition}, \rf{faction} $\Psi_f(x)$ is the Grassmann
anticommuting $N_c\times N_c$ matrix field while
\be
P_\mu^\pm=r\pm\gamma_\mu \label{projectors}
\ee
stand for the projectors. The case $r=0$ corresponds to chiral fermions
while $r=1$ is associated with Wilson fermions. As is well
known, the chiral fermions describe $2^D N_f$ flavors in the naive continuum
limit while Wilson fermions are associated with $N_f$ flavors.

\subsection{The induced action}

The above matrix models can be represented in the form of a
{\it gauge theory\/} given by the partition function
\be
Z=\int \prod_{x,\mu} dU_\mu(x)\,\e^{-S_{ind}[U_\mu(x)]}\;,
 \label{gaugetheory}
 \ee
 where the {\it induced action\/} for the gauge field $U_\mu(x)$ is defined
 by the integral over $\Phi(x)$ in \eq{spartition} or over $\Psi(x)$ and
 $\bar{\Psi}(x)$ in \eq{fpartition}:
\be
\e^{-S_{ind}[U_\mu(x)]}=\int \prod_x \prod_{f=1}^{N_f} d\Phi_f(x)
\e^{\sum_{f=1}^{N_f} \sum_x
N \tr{\left(-V[\Phi_f(x)]+
\sum_{\mu=1}^D\Phi_f(x)U_\mu(x)\Phi_f(x+\mu)U_\mu^\dagger(x)\right)}}
\label{sinduced}
\ee
or
\be
\e^{-S_{ind}[U_\mu(x)]}=\int \prod_x \prod_{f=1}^{N_f} d\Psi_f(x)
d\bar{\Psi}_f(x) \e^{-S_{F}[\Psi,\bar{\Psi},U]} \,,
\label{finduced}
\ee
with $S_{F}[\Psi,\bar{\Psi},U]$ given by \eq{faction}.

For the quadratic potential the result
of integrating over $\Phi(x)$ or over $\Psi(x)$ and $\bar{\Psi}(x)$
is given by the large mass expansion:
\be
S_{ind}[U] = - \frac{N_f}{2} \sum_{\Gamma}
\frac{|\tr{U(\Gamma)}|^2}{l(\Gamma)}\left(\frac{2}{m}\right)^{l(\Gamma)}
\label{sind}
\ee
for scalars or
\be
S_{ind}[U] = - N_f \sum_{\Gamma}
\frac{|\tr{U(\Gamma)}|^2}{l(\Gamma)m^{l(\Gamma)}}\, \hbox{Sp}\,
\prod_{l\in \Gamma} P_\mu^\pm
\label{find}
\ee
for fermions. In \eq{find}
\ Sp \ stands for the trace over the spinor indices of the
path-ordered product of the projectors \rf{projectors} (plus or minus
depends on the orientation of the link $l$) along the loop $\Gamma$.

One easily sees that only single plaquettes survive in
the sum over path on the r.h.s.'s of Eqs.~\rf{sind} and \rf{find} if
 $m\sim N_f^{1/4}$ as $N_f\ra \infty$, so that the single
plaquette adjoint action arises in the large-$N_f$ limit:
\be
S_{A}= -
\frac{\beta_A}{2} \sum_p \left|\tr{U(\d p)}\right|^2 \label{adjaction}
\ee
with
\bea
\beta_A= \frac{4N_f}{m^4}& & \hbox{ \ \ \ \ \ \ for scalars};
\nonumber \\* \beta_A= \frac{2^{\fr D2 -1}N_f(1+2r^2-r^4)}{m^4} & & \hbox{
\ \ \ \ \ \ for fermions}\,.
\eea
This shows of how ASM and AFM induce the
single plaquette lattice gauge theory with adjoint action.

\subsection{The large-$N$ phase transition}

The inducing of large-$N_c$ QCD relies on the fact~\cite{KhM81} that the
lattice gauge theory defined by the partition function~\rf{gaugetheory} with
the action \rf{adjaction} undergoes the first order large-$N_c$
phase transition at $\beta_A\approx 2$ after which the gauge field
$U_\mu(x)$ becomes frozen near some mean-field value $\h$ ($\h\ra1$ as
$\beta_A\ra\infty$).

The proof of the existence of the phase transition is based solely on the
factorization at large-$N_c$ which says that the adjoint
action~\rf{adjaction} is equivalent at $N_c=\infty$ to the single-plaquette
{\it fundamental\/} action
\be
S_{F}[U]=
N_c \bb \sum_p \Re \tr{}U(\d p) \label{fundaction}
\ee
providing the coupling $\bb$ is determined by
\be
\bb=\beta_A W_F (\d p;\bb)
\label{bb}
\ee
where $W_F (\d p;\bb)$ stands for the plaquette average
\be
W_F (\d p;\bb) \equiv \frac {\int \prod_{x,\mu} dU_\mu(x)\,
\e^{-S_F[U]}\frac {1}{N_c}\tr{U(\d p)}}{\int
\prod_{x,\mu} dU_\mu(x)\,\e^{-S_F[U]}}\,.
\label{plaquetteaverage}
\ee
\eq{bb} can be naively obtained substituting one of two traces in the
action~\rf{adjaction} by the average. A rigorous proof~\cite{KhM81} is based on
the loop equations.

The existence of the first order phase transition for the
action~\rf{adjaction} with decreasing $\beta_A$ can be seen as follows.
Let us solve \eq{bb} for $\bb(\beta_A)$ substituting
for~\rf{plaquetteaverage} the strong coupling expansion
\be
W_F (\d p;\bb) = \frac \bb2+ \frac {\bb^5}{8}\hspace{2cm}\hbox{at strong
coupling}\,.
\label{strongcoupling}
\ee
\eq{bb} possesses at any $\bb$ a trivial solution $\bb=0$. However,
one more solution appears for $\bb\approx2$:
\be
\bb\propto\Big( \frac 12-\frac{1}{\beta_A} \Big)^{\frac 14}\,,
\label{newsolution}
\ee
which matches the weak coupling solution
\be
\bb\ra\beta_A-\frac 14 \;\;\;\hbox{as}\;\;\;\beta_A\ra\infty\,.
\ee
Notice that $\bb\ll1$ for the solution~\rf{newsolution} when
$\beta_A\approx2$ so that the strong coupling expansion is applicable.

The adjoint plaquette average
\be
W_A (\d p;\beta_A) \equiv \frac {\int \prod_{x,\mu} dU_\mu(x)\,
\e^{-S_A[U]}\frac {1}{N_c^2}\Big(\left|\tr{U(\d p)}\right|^2-1\Big)
}{\int \prod_{x,\mu} dU_\mu(x)\,\e^{-S_A[U]}}
\label{aplaquetteaverage}
\ee
which is given due to the factorization by
\be
W_A (\d p;\beta_A) =\Big( W_F (\d p;\bb)\Big)^2=
\left(\frac{\bb}{\beta_A}\right)^2
\label{factorization}
\ee
is depicted in Fig.~$1$. Since the slope is negative for
the solution~\rf{newsolution} near $\beta_A=2$, a first order phase
transition must occur with increasing $\beta_A$. This negative slop is a
consequence solely of the positive sign of the second term on the r.h.s.\ of
\eq{strongcoupling}. The predicted value of the critical coupling $\beta_A^*$,
at which the phase transition occurs, obeys $\beta_A^*<2$, as is seen from
Fig.~$1$, to be compared with the result of Monte--Carlo simulations
$\beta_A^*=1.7 - 1.8$.

\subsection{The mean field phase diagram}

At finite $N_f$ the induced actions~\rf{sind} or \rf{find} can not be exactly
analyzed even at $N_c=\infty$. An approximate mean field method, which usually
works very well in the cases of first order phase transitions,
was applied to obtain the phase diagram in Refs.~\cite{KhM92a,KhM92b}.

Naively, the mean field approximation consists in substituting the link
variable $U_\mu(x)$ by the mean field value
\be
 [U_\mu(x)]^{ij}=\h\, \delta^{ij}
\ee
everywhere but one link and writing a self-consistency condition at this link.
The self-consistency condition is given by the one-link problem
\be
\h^2 = \frac{\int dU
\e^{\frac{b_A}{2}\left|\tr{U} \right|^2} \frac{1}{N^2}\left|\tr{U} \right|^2 }
{\int dU \e^{\frac{b_A}{2}\left|\tr{U} \right|^2}}
\label{mf}
\ee
where
\be
b_A= \frac{\int \prod_{x}d\Phi(x)
\e^{\sum_x N\tr{\left(-V[\Phi(x)]
+\h^2\sum_\mu \Phi(x)\Phi(x+\mu)\right)}}
\fr 1N \tr{\Phi(0)}\Phi(0+\mu)}
{\int \prod_{x}d\Phi(x)
\e^{\sum_x N\tr{\left(-V[\Phi(x)]
+\h^2\sum_\mu \Phi(x)\Phi(x+\mu)\right)}}}\;.
\label{b_A}
\ee
These naive mean field formulas can be obtained~\cite{KhM92a} in the framework
of the variational method.

The mean field phase diagram which was obtained by an analysis of Eqs.~\rf{mf}
and \rf{b_A} is depicted in Fig.~$2$. At $N_f=1$ there is no first order phase
transition for the quadratic potential in the stability region $m>2D$. For
$m<2D$ the model is unstable and were in the Higgs phase if the stabilizing
higher order in $\Phi$ terms would be added to the potential~\rf{potential}.
The desired large-$N_c$ phase transition appears for $N_f>N_f^* \approx 30$.%
\footnote{Such a phase diagram is compatible with the Monte--Carlo studies of
Ref.~\cite{GS92}.}
ASM looks in this region exactly like the single-plaquette
adjoint model discussed in the previous subsection.

A similar phase diagram for AFM is depicted in Fig.~$3$.
Now there is no Higgs phase (or an unstability region for the quadratic
potential) due to the fermionic nature of inducing fields. For the cases of
chiral and Kogut--Susskind fermions the first order phase transition is
 present already for $N_f=1$ while the result for Wilson fermions is
less certain.

\subsection{Area law versus local confinement}

At the point of the first order large-$N_c$ phase transition, the area law
behavior of the (adjoint) Wilson loops which is associated with normal
confinement is restored in ASM or AFM analogously to the single-plaquette
adjoint action~\cite{KhM81}.

In order to see this, let us consider the adjoint
Wilson loop which is defined by
\be
W_A (C) =
\left\langle \frac {1}{N_c^2} \Big(\left|\tr{U(C)}\right|^2-1\Big)
\right\rangle
\label{adjloop}
\ee
where the average is understood w.r.t.\ the same measure as in \eq{spartition}
or in \eq{fpartition}. Alternatively, one can average w.r.t.\ the induced
actions~\rf{sind} \rf{find} which recovers at $N_f=\infty$ the single plaquette
adjoint action~\rf{adjaction}.
In this limiting case the following extension of the factorization
formula~\rf{factorization} holds at $N_c=\infty$:
\be
W_A (C;\beta_A) =\Big( W_F (C;\bb)\Big)^2\,,
\label{Cfactorization}
\ee
where $W_F(C;\bb)$ is defined by the same formula as~\rf{plaquetteaverage}
with $\d p$ replaced by an arbitrary contour $C$ and $\bb$ versus $\beta_A$
given by \eq{bb}.

Since $\bb=0$ for $\beta_A<\beta_A^*$, $W_A(C)$ vanishes
in this region due to \eq{Cfactorization} except the loops with vanishing
minimal area $A_{min}(C)$:
\be
W_A(C)=\delta_{0A_{min}(C)}+{\cal O}\left({1\over N_c^2} \right)\,.
\label{1overN}
\ee
On the contrary, the area law with the string tension
\be
K_A(\beta_A)=2\,K_F(\bb(\beta_A))
\ee
holds for $\beta_A>\beta_A^*$ when \eq{bb} possesses the non-trivial solution.
An extension of these formulas to finite $N_f$ is given in Ref.~\cite{Mak92}.

While the first order phase transition associated with the restoration of area
law looks similar for ASM and AFM, the continuum limits should be approached in
different ways. For ASM the continuum QCD is reached at the line of second
order phase transitions which separates the area law and Higgs phases provided
that one approaches it from the area law phase.  For AFM there is no Higgs
phase and continuum QCD is reached as $m\ra0$.

\newsection{Loop equations at strong coupling \label{loops}}

The loop equations of ASM or AFM relate the closed adjoint Wilson
loop~\rf{adjloop} to the open ones with scalars or fermions at the ends.
The loop equations are drastically simplified at $N_c=\infty$ in the strong
coupling region where the closed loops obey \eq{1overN}. The exact solution can
be obtained in both cases for the quadratic potential when the loop equations
turns out to be equivalent to those for the hermitean and complex one-matrix
models, respectively.

\subsection{Loop equations for arbitrary potential}

The generic object which appear in the loop equations are open Wilson loops
\be
\delta_{ff^\p}G_{\l}(C_{xy})= \left\langle
\frac {1}{N_c} \tr{\left( \Phi_f(x) U (C_{xy})
\frac{1}{\l- \Phi_{f^\p}(y)}
U^\dagger(C_{xy}) \right)} \right\rangle\,.
\label{sG}
\ee
The appearance of the $\delta$-symbol w.r.t.\ the flavor indices $f$ and
$f^\p$ is due to the fact that the interaction terms in the action
entering \eq{spartition} are diagonal over the flavor indices.

The loop equations of ASM result from the invariance of the measure in
\eq{spartition} under an arbitrary shift of $\Phi_f(x)$ and reads
\bea
 \left\langle \frac{1}{N_c} \tr{}\Big(V^\p(\Phi_f(x))
U(C_{xy}) \frac{1}{\l-\Phi_{f^\p}(y)}
U^\dagger(C_{xy}) \Big) \right\rangle
\nonumber \\* -\sum_{\mu=-D\atop \mu\neq0}^D
\left\langle \frac{1}{N_c} \tr{\Big( \Phi_{f}(x+\mu)
U(C_{(x+\mu)x}C_{xy})
 \frac{1}{\l-\Phi_{f^\p}(y)}U^\dagger(C_{(x+\mu)x}C_{xy})}\Big)
 \right\rangle = \nonumber \\* \cdot  \delta_{ff^\p} \delta_{xy}
\left\langle \frac{1}{N_c} \tr{}{\Big( U(C_{xy})\frac{1}{\l-\Phi_{f}(y)}\Big)}
\frac{1}{N_c} \tr{}{ \Big(\frac{1}{\l-\Phi_{f}(y)}
U^\dagger(C_{xy})\Big)}\right\rangle
\label{sd}
\eea
where the path $C_{(x+\mu)x}C_{xy}$
on the l.h.s.\ is obtained by attaching the link $(x,\mu)$ to the path
$C_{xx}$ at the end point $x$ as is depicted in Fig.~4. I have omitted
additional contact
terms which arise at finite $N_c$ due to the fact that $\Phi$ belongs to
the adjoint representation, so that \eq{sd} is written for the
hermitean matrices. This difference disappears, however, as $N_c\ra\infty$.

The analogues of Eqs.~\rf{sG} and \rf{sd} for AFM read
\be
\delta_{ff^\prime}{G}^{ij}_{\l}(C_{xy})=
\left\langle \frac{1}{N_c}\, \hbox{tr}
\Big(  \Psi^i_f(x) U (C_{xy})
\frac{\l \bar{\Psi}^j_{f^\prime}(y)}
{\l^2-\bar{\Psi}_{f^\prime}(y)\Psi_{f^\prime}(y)}
U^\dagger(C_{xy}) \Big)
 \right\rangle
\label{fG}
\ee
where $i$ and $j$ are spinor indices, and \bea \Big\langle \frac{1}{N_c}
 \tr{\Big( \Psi_f(x)V^\p_{even}({\bar{\Psi}_f(x)\Psi_f(x)}) U(C_{xy})
\frac{\bar{\Psi}_f(x)}{\l-\bar{\Psi}_f(x)\Psi_{f^\p}(y)}
U^\dagger(C_{xy})\Big)} \Big\rangle  \nonumber \\*
-\sum_{\mu=1}^D \Big\langle \frac{1}{N_c} \tr{}{\Big(P_\mu^+ \Psi_{f}(x+\mu)
U(C_{(x+\mu)x}C_{xy})
 \frac{\l\bar{\Psi}_{f^\p}(y)}{\l-\bar{\Psi}_{f^\p}(y)\Psi_{f^\p}(y)}
 U^\dagger(C_{(x+\mu)x}C_{xy})} \nonumber \\
 +P_\mu^- \Psi_{f}(x-\mu)U(C_{(x-\mu)x}C_{xy})
 \frac{\l\bar{\Psi}_{f^\p}(y)}{\l-\bar{\Psi}_{f^\p}(y)\Psi_{f^\p}(y)}
 U^\dagger(C_{(x-\mu)x}C_{xy}) \Big)
\Big\rangle \nonumber \\ = \delta_{ff^\p} \delta_{xy}
\Big\langle \frac{1}{N_c} \tr{ \Big(U(C_{xy})\frac{\l}{\l^2-
\bar{\Psi}_{f}(y)\Psi_{f}(y)}\Big)}
\frac{1}{N_c} \tr{\Big( \frac{\l}{\l^2-\bar{\Psi}_{f}(y)\Psi_{f}(y)}
U^\dagger(C_{xy})\Big)}
\Big\rangle\,.
\label{fsd}
\eea
The matrix multiplication over the spinor indices is implied in this equation.

\subsection{Loop equations at large $N_c$}

The path $C_{xy}$ on the r.h.s.\ of \eq{sd} (or \eq{fsd}) is always
closed due to the presence of the delta-function. The explicit equation for
the case of vanishing (or contractable) contour $C_{xx}=0$ at large $N_c$, when
the factorization holds, reads
\be
\int_{C_1}\frac {d\omega}{2\pi i}
\frac{V^\p(\omega)}{(\l-\omega)} E_{\omega}
-2D G_{\l}(1) =E_\l^2
\label{sd0}
\ee
where
\be
E_\l \equiv \left\langle
\frac{1}{N_c}\tr{}\Big( \frac{1}{\l-\Phi_f(x)} \Big) \right\rangle
=\frac{1}{\l}( G_\l(0)+1)
\label{defE}
\ee
with $G_\l$ is defined by \eq{sG}.
I have denoted the one-link average by
\be
G_{\l}(1)=G_{\l}(C_{(x\pm\mu)x})
\ee
since the r.h.s.\ does not depend on $x$ and $\mu$ due to the
invariance under translations by a multiple of the lattice spacing and/or
rotations by a multiple of $\pi/2$ on the lattice.
The contour $C_1$ encircles singularities of ${E}_{\omega}$ so that the
integration over $\omega$ on the l.h.s.\ of \eq{sd0} plays the role of a
projector picking up negative powers of $\l$.

For $C_{xx}\neq0$, the averages of a new kind
arise on the r.h.s.\ of \eq{sd} (or \eq{fsd}). However, these averages obey at
$N_c=\infty$ the following analogue of \eq{1overN}
\be
\left\langle
\frac{1}{N_c}\tr{}{\Big(U(C_{xx})\frac{1}{\l-\Phi_f(x)}\Big)}
\frac{1}{N_c}\tr{}{\Big(U^\dagger(C_{xx})
\frac{1}{\l-\Phi_f(x)}\Big)}\right\rangle  =
\delta_{0,A_{min}(C)} E_\l^2
+{\cal O}\left({1\over N_c^2} \right)
\label{1overNp}
\ee
\ie vanish for $C_{xx}\neq0$.

Hence, the loop equation for $C_{xy}\neq0$ at $N_c=\infty$ reads
\bea
 \left\langle \frac{1}{N_c} \tr{}\Big(V^\p(\Phi_f(x))
U(C_{xy}) \frac{1}{\l-\Phi_{f^\p}(y)}
U^\dagger(C_{xy}) \Big) \right\rangle
-\sum_{\mu=-D\atop \mu\neq0}^D G_\l(C_{(x+\mu)x}C_{xy})=0
\label{sd1}
\eea
independently of whether $C_{xy}$ is closed or open.
 Therefore,
the r.h.s.\ of the loop equation in nonvanishing at $N_c=\infty$ only for
$C_{xy}=0$ (modulo backtrackings) when the proper equation is given
by \eq{sd0}.

Finally, the fermionic analogues of Eqs.~\rf{1overNp} and \rf{defE} read
\bea
\left\langle
\frac{1}{N_c}\tr{}\left(U(C_{xx})\frac{\l}{\l^2-\bar{\Psi}_f(x)\Psi_f(x)}\right)
\frac{1}{N_c}\tr{}{\left(U^\dagger(C_{xx})\frac{\l}{\l^2-
\bar{\Psi}_f(x)\Psi_f(x)}\right)}\right\rangle \nonumber \\* =
\delta_{0,A_{min}(C)}E_\l^2+{\cal O}\left({1\over N^2} \right)
\eea
and
\be
E_\l =
\left\langle
\frac{1}{N_c}\tr{}\Big(\frac{\l}{\l^2-\bar{\Psi}_f(x)\Psi_f(x)}\Big)
 \right\rangle
=\frac{1}{\l}( G_\l^{ii}(0)+1)
\ee
with $G_\l$ defined by \eq{fG}.
Therefore, the r.h.s.\ of \eq{fsd} involves at $N_c=\infty$ only $E_\l$
similarly to the scalar case.

\subsection{The quadratic potential}

The quadratic potential is always solvable, even in the
non-diagonizable cases, for the following reasons. Let us consider
the one-link correlator
\be
\Big\langle \frac {1}{N_c} \tr{} t^aU\chi_{f^\p} U^\dagger \Big\rangle_U
\equiv\frac{\int dU\,\e^{N_c\sum_f \varphi_f U \chi_f U^\dagger}
\frac {1}{N_c} \tr{} t^aU\chi_{f^\p} U^\dagger }
{\int dU\,\e^{N_c\sum_f \varphi_f U \chi_f U^\dagger}}
\label{onelink}
\ee
where $t_a$ ($a=1,\ldots,N_c^2-1$) stand for generators of $SU(N_c)$ which are
normalized by
\be
\frac{1}{N_c}\tr{}t^at^b = \delta^{ab} \,.
\ee
At $N_c=\infty$ the formula
\be
\Big\langle \frac {1}{N_c} \tr{} t^aU\chi_{f^\p} U^\dagger \Big\rangle_U
=\Lambda \frac{1}{N_c}\tr{}t^a\varphi_{f^\p}\;,
\label{Lambda}
\ee
where $\Lambda$ is a constant to be determined below,
can be proven for $\varphi$ and $\chi$ given by the master field for the
quadratic potential analyzing the large mass expansion.
The point is that terms like $\tr{}t^a\varphi^k_{f^\p}$ with $k>1$ never
appear for the quadratic potential.
Analogously it can be proven that
\be
\Big\langle \frac {1}{N_c} \tr{} t^a U\frac{1}{\l-\chi_{f^\p}} U^\dagger
\Big\rangle_U =\Lambda \frac{1}{N_c}\tr{}t^a\frac{1}
{\l- \varphi_{f^\p}}\,.
\label{LLambda}
\ee
For $N_f=1$ Eqs.~\rf{Lambda}  and \rf{LLambda} recovers the ones of
Ref.~\cite{Mig92d}.

For $G_\l(C_{xy})$ defined by \eq{sG}, \eq{LLambda} implies
\be
G(C_{xy}) = \Lambda^L E_\l
\label{ansatz}
\ee
where $L$ is the {\it algebraic\/} length (\ie the one
after contracting the backtrackings) of $C_{xy}$.
The fermionic analogue of this formula reads
\be
G_{\l}^{ij}(C_{xy})= \Lambda^L E_\l \left(\prod_{l\in
C_{xy}}P_\mu^\pm \right)^{ij}
\label{spinfactor}
\ee
where the plus or minus signs
correspond to the direction of the link $l$ which belongs to the contour
$C_{xy}$. The spin factor will provide below the
cancellation of the projectors in Eq.~\rf{fsd}.

The constant $\Lambda$ can be determined by substituting the
ansatz~\rf{ansatz} into the $C_{xy}\neq0$ loop equation~\rf{sd1} which
simplifies for the quadratic potential as~\cite{Mak92}
\be
m G_\l(C_{xy})-\sum_{\mu=-D\atop \mu\neq0}^D G_\l(C_{(x+\mu)x}C_{xy})=0.
\label{qsd1}
\ee
The ansatz~\rf{ansatz} satisfies this equation for any
$C_{xy}\neq0$ providing
\be
\Lambda=\frac{2}{m+\sqrt{m^2+4(1-2D)\sigma}}
\label{Lphi}
\ee
where $\sigma=1$ for scalars or
\be
\sigma=P_\mu^+ P_\mu^- =r^2-1
\label{sigma}
\ee
 for fermions ($\sigma=-1$ for chiral fermions and $\sigma=0$ for
Wilson fermions).

The remaining function $E_\l$ can now be determined from \eq{sd0} which for the
quadratic potential reads
\be
\tilde m  E_\l = E_\l^2\,, \;\;\;\;
\tilde m = m-2D\Lambda
\label{le}
\ee
and coincides with the loop equation for the Gaussian hermitean one-matrix
model (for a review, see Ref.~\cite{Mak91} and references therein). The
solution of \eq{le} which satisfies
\be
E_\l\ra \frac 1\l \;\;\;\; \hbox{ as \ } \l\ra\infty \;,
\ee
as it should be due to the definition~\rf{defE}, is unambiguous:
\be
2 E_\l =  \tilde m \l - \tilde m \sqrt{\l^2-4/\tilde m}\,.
\label{quadrsolution}
\ee
The imaginary part
\be
\Im E_\l\equiv \rho(\l) = \frac{1}{4\pi} \tilde m \sqrt{4/\tilde m-\l^2}
\;\;\;\; \hbox{ for \ \ } -2/\sqrt{\tilde m}\leq \l \leq 2/\sqrt{\tilde m}
\ee
recovers the solution~\cite{Gro92}. Analogously, the $C_{xx}=0$ loop equation
for AFM with the quadratic potential is reduced to the loop equation for the
complex one-matrix model~\cite{Mak91}.

One should not be surprised that the exact strong coupling solution for the
quadratic potential does not depend on $N_f$ which is a consequence of the
peculiar behavior of the Wilson loops~\rf{1overN}. This independence does not
contradict to the fact that the first order phase transition discussed in
Sect.~\ref{Mamo} occurs only for $N_f>N_f^*$. The point is that the strong
coupling solution is not sensitive to the phase transition which occurs due to
another (thermodynamic) reason.

\subsection{Interpretation as the $1D$ tree problem}

A question arises what combinatorial problem are the exact solutions of
the previous subsection associated with? To answer, let us consider the open
loop correlator
\be
\delta_{ff^\p}G(C_{xy})= \left\langle
\frac {1}{N_c} \tr{\left( \Phi_f(x) U (C_{xy})
\Phi_{f^\p}(y)
U^\dagger(C_{xy}) \right)} \right\rangle\,.
\label{sGp}
\ee
which is nothing but the $\l^{-2}$ term of the expansion of~\rf{sG} in
$\l^{-1}$. At $N_c=\infty$ the standard sum-over-path representation
of $G(C_{xy})$ reads
\be
G(C_{xy}) =\sum_{\Gamma_{yx}} \left(\frac{2}{m}\right)^{l(\Gamma)+1}
W_A(C_{xy}\Gamma_{yx})
\label{path}
\ee
where the contour $\Gamma_{yx}$ forms together with $C_{cy}$ a
closed loop passing $x$ and $y$.

Since in our case the formula~\rf{1overN} associated with the infinite string
tension holds, $\Gamma_{yx}$ must coincide with $(C_{xy})^{-1}$ (\ie passed in
opposite direction) modulo backtrackings of $\Gamma$. \eq{path} then yields
\be
G(C_{xy}) =\sum_{\Gamma_{yx}} \left(\frac{2}{m}\right)^{l(\Gamma)+1}
\label{path0}
\ee
where the sum goes over contours of the type depicted in Fig.~$5$.%
\footnote{One should not confuse the double lines in Fig.~$5$ which
are due to backtrackings with the double lines in Fig.~$4$ which represent the
adjoint representation.}

The fermonic analogues of Eqs.~\rf{sGp} to \rf{path0} read
\be
G^{ij}(C_{xy}) =\sum_{\Gamma_{yx}} \left(\frac{1}{m}\right)^{l(\Gamma)+1}
W_A(C_{xy}\Gamma_{yx})
\left(\prod_{l\in C_{xy}}P_\mu^\pm \right)^{ij}
\label{fpath}
\ee
and
\be
G^{ii}(C_{xy}) =\sum_{\Gamma_{yx}} \left(\frac{1}{m}\right)^{l(\Gamma)+1}
\, \hbox{Sp}\,\left(\prod_{l\in C_{xy}}P_\mu^\pm \right)
\label{fpath0}
\ee
with summing again over contours depicted in Fig.~$5$.

The proper combinatorial
problem is, therefore, that of summing over $1$-dimensional trees embedded in a
$D$-dimensional space. The exact solution of loop equations for the quadratic
potential represents the solution to this problem:
\be
G(C_{xy})= \Lambda^L \frac{2D-1}{m(D-1)+D\sqrt{m^2+4(1-2D)\sigma}}\,,
\label{solut}
\ee
where $L$ is the algebraic length of $C_{xy}$ and $\Lambda$ is defined by
\eq{Lphi}. Such a dependence on $\Lambda$ is evident from the
representation~\rf{path0} (or \rf{fpath0}) since the trees are uniformly
distributed along $C_{xy}$.

As is already mentioned in the previous section, this solution coincides for
scalars with that of Ref.~\cite{Gro92}. \eq{solut} takes an especially simple
form for Wilson fermions when the backtracking parameter $\sigma$, given by
\eq{sigma}, vanishes so that there are no backtrackings. For chiral fermions
when $\sigma=-1$ the solution~\rf{solut} coincides with that of
Ref.~\cite{KMNP81} for the case of the fundamental representation and vanishing
constant in front of the plaquette term in the action. The point is that Wilson
loops vanish in this case as well (except for those with vanishing minimal
area) and exactly the same combinatorial problem of summing the diagrams
of the type depicted in Fig.~$5$ emerges.

\section*{Acknowledgement}

I am grateful to the theoretical physics department of the University of
Zaragoza for the hospitality last December when the manuscript
was being prepared for publication.

\eop

\eop
\section*{Figures}

\vspace{3cm}
\hspace{0.4cm}
\unitlength=1.00mm
\linethickness{0.5pt}
\begin{picture}(120.00,110.00)
\put(40.00,110.00){\line(1,0){80.00}}
\put(40.00,110.00){\line(0,-1){50.00}}
\put(120.00,110.00){\line(0,-1){50.00}}
\put(40.00,60.00){\line(1,0){80.00}}
\put(75.00,66.50){\oval(10.00,13.00)[lb]}
\put(75.00,66.50){\oval(10.00,10.00)[lt]}
\thicklines
\put(73.00,71.10){\line(6,1){46.50}}
\put(72.00,60.00){\line(0,1){10.50}}
\put(40.00,60.00){\line(1,0){32.00}}
\thinlines
\put(35.00,79.00){\makebox(0,0)[cc]{$1$}}
\put(38.00,79.00){\line(1,0){4.00}}
\put(75.00,55.00){\makebox(0,0)[lc]{$2$}}
\put(72.00,60.00){\line(0,-1){2.00}}
\put(72.00,55.00){\makebox(0,0)[rc]{$\beta_A^*$}}
\put(74.50,60.00){\line(0,-1){2.00}}
\put(45.00,100.00){\makebox(0,0)[lc]{$W_A(\d p; \beta_A)$}}
\put(80.00,45.00){\makebox(0,0)[cc]{$\beta_A$}}
\put(40.00,55.00){\makebox(0,0)[lc]{$0$}}
\put(120.00,55.00){\makebox(0,0)[cc]{$\infty$}}
\put(35.00,60.00){\makebox(0,0)[cb]{$0$}}

\end{picture}
\vspace{-3.5cm}
\begin{description}
\item[Fig. 1] {\small
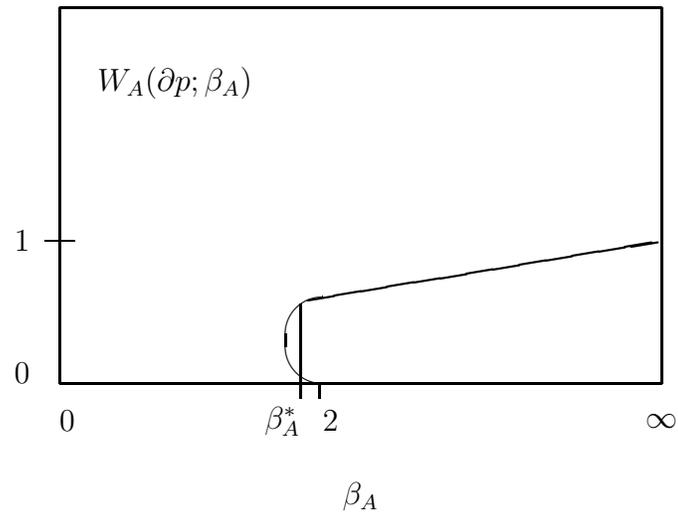
\ \ \ \ The two solutions of \eq{bb}. The line which starts at $\beta_A=2$
is associated with the solution~\rf{newsolution}. Since the slope is negative
for this solution near $\beta_A=2 $ (and $W_A\ll1$), the first order phase
transition occurs at some $\beta_A^*<2$ so that the actual behavior of $W_A(\d
p;\beta_A)$ is depicted by the bold line. } \end{description}

\eop
\vspace*{-0.1cm}

\unitlength=1mm
\linethickness{0.5pt}
\begin{picture}(120.00,110.00)
\put(40.00,110.00){\line(1,0){80.00}}
\put(40.00,110.00){\line(0,-1){50.00}}
\put(120.00,110.00){\line(0,-1){50.00}}
\put(40.00,60.00){\line(1,0){80.00}}
\thicklines
\put(75.00,60.00){\line(0,1){20.00}}
\put(75.00,80.00){\line(-2,3){20.00}}
\thinlines
\put(75.00,80.00){\line(2,3){20.00}}
\put(35.00,80.00){\makebox(0,0)[cc]{$N_f^*$}}
\put(35.00,110.00){\makebox(0,0)[cc]{$\infty$}}
\put(30.00,85.00){\makebox(0,0)[rc]{$N_f$}}
\put(38.00,80.00){\line(1,0){4.00}}
\put(75.00,55.00){\makebox(0,0)[cc]{${1}/{2D}$}}
\put(80.00,45.00){\makebox(0,0)[cc]{$1/m$}}
\put(40.00,55.00){\makebox(0,0)[lc]{$0$}}
\put(120.00,55.00){\makebox(0,0)[cc]{$\infty$}}
\put(35.00,60.00){\makebox(0,0)[cb]{$1$}}
\put(75.00,102.00){\makebox(0,0)[cc]{Area}}
\put(75.00,97.00){\makebox(0,0)[cc]{law}}
\put(57.00,76.00){\makebox(0,0)[cc]{Local}}
\put(57.00,71.00){\makebox(0,0)[cc]{confinement}}
\put(96.00,76.00){\makebox(0,0)[cc]{Higgs}}
\put(96.00,71.00){\makebox(0,0)[cc]{phase}}
\end{picture}

\vspace{-4cm}
\begin{description} {\small
\item[Fig. 2]
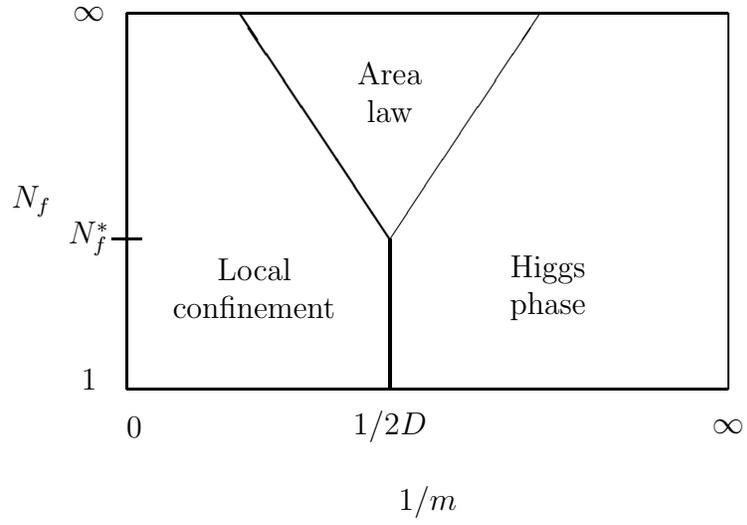
\ \ \ \ The mean field prediction for the phase diagram of ASM.
The bold lines which bounds the local confinement phase is that of
first order phase transitions. The line which separates the area law and
Higgs phases is that of second order phase transitions. }
\end{description}

\vspace{2cm}
\unitlength=1.00mm
\linethickness{0.5pt}
\begin{picture}(120.00,110.00)
\put(40.00,110.00){\line(1,0){80.00}}
\put(40.00,110.00){\line(0,-1){50.00}}
\put(120.00,110.00){\line(0,-1){50.00}}
\put(40.00,60.00){\line(1,0){80.00}}
\thicklines
\put(75.00,60.00){\line(0,1){50.00}}
\thinlines
\put(35.00,110.00){\makebox(0,0)[cc]{$\infty$}}
\put(30.00,85.00){\makebox(0,0)[rc]{$N_f$}}
\put(75.00,55.00){\makebox(0,0)[cc]{$1/m^*$}}
\put(80.00,45.00){\makebox(0,0)[cc]{$1/m$}}
\put(40.00,55.00){\makebox(0,0)[lc]{$0$}}
\put(120.00,55.00){\makebox(0,0)[cc]{$\infty$}}
\put(35.00,60.00){\makebox(0,0)[cb]{$1$}}
\put(57.00,86.00){\makebox(0,0)[cc]{Local}}
\put(57.00,81.00){\makebox(0,0)[cc]{confinement}}
\put(96.00,86.00){\makebox(0,0)[cc]{Area}}
\put(96.00,81.00){\makebox(0,0)[cc]{law}}
\end{picture}

\vspace{-4cm}
\begin{description} {\small
\item[Fig. 3]
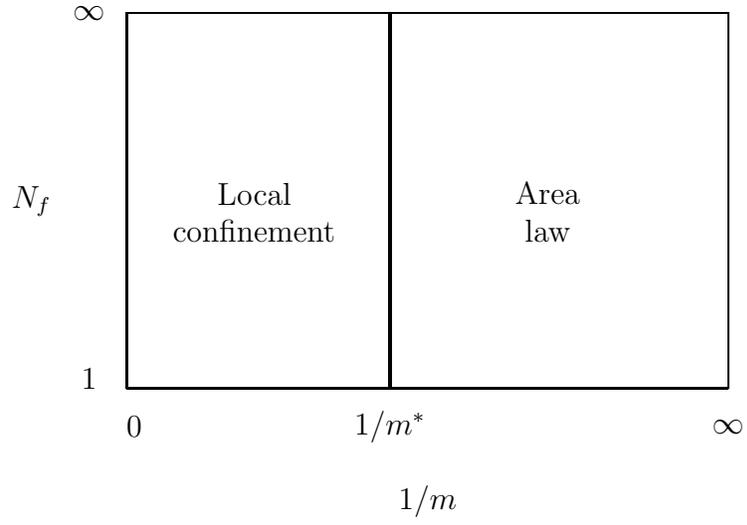
\ \ \ \ The mean field prediction for the phase diagram of AFM.
The bold line which separates the local
confinement and the area law phases is that of first order phase
transitions.}  \end{description}

\eop
\vspace*{-0.2cm}
\hspace{1.4cm}
\unitlength=0.70mm
\linethickness{0.5pt}
\begin{picture}(141.00,90.00)
\put(30.00,50.00){\line(3,0){20.00}}
\put(50.00,90.00){\line(3,0){20.00}}
\put(100.00,50.00){\line(3,0){20.00}}
\put(120.00,90.00){\line(3,0){20.00}}
\put(30.00,55.00){\makebox(0,0)[cc]{$x$}}
\put(70.00,83.00){\makebox(0,0)[cc]{$y$}}
\put(100.00,55.00){\makebox(0,0)[cc]{$x$}}
\put(140.00,83.00){\makebox(0,0)[cc]{$y$}}
\put(90.00,44.00){\line(5,3){10.00}}
\put(90.00,37.00){\makebox(0,0)[cc]{$x+\mu$}}
\put(50.00,20.00){\makebox(0,0)[cc]{a)}}
\put(120.00,20.00){\makebox(0,0)[cc]{b)}}
\put(50.00,50.00){\vector(0,1){20.00}}
\put(50.00,90.00){\line(0,-3){20.00}}
\put(120.00,50.00){\vector(0,1){20.00}}
\put(120.00,70.00){\line(0,3){20.00}}
\put(30.00,49.00){\circle{2.00}}
\put(70.00,89.00){\circle*{2.00}}
\put(52.00,88.00){\vector(0,-1){20.00}}
\put(52.00,48.00){\line(0,3){20.00}}
\put(30.00,48.00){\line(3,0){22.00}}
\put(52.00,88.00){\line(3,0){18.00}}
\put(90.00,43.00){\circle{2.00}}
\put(140.00,89.00){\circle*{2.00}}
\put(122.00,88.00){\line(3,0){18.00}}
\put(122.00,88.00){\vector(0,-1){20.00}}
\put(122.00,48.00){\line(0,3){20.00}}
\put(100.00,48.00){\line(3,0){22.00}}
\put(90.00,42.00){\line(5,3){10.00}}
\end{picture}

\vspace{-0.5cm}

\begin{description} {\small
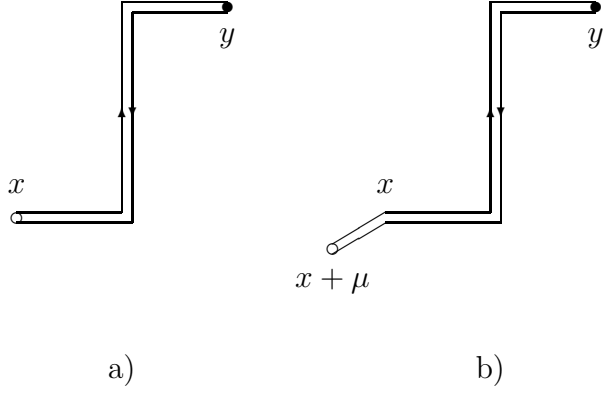
   \item[Fig. 4] \ \ \ The graphic representation for $G_\l(C_{xy})$ (a)
   and $G_\l(C_{(x+\mu)x}C_{xy})$ (b) entering \eq{sd}. The
   empty circles represent ${\Phi}_f(x)$ or $\Phi_f(x+\mu)$ while the filled
   ones represent $\frac{1}{\l-\Phi_{f^\p}(y)}$.
   The oriented solid lines represent the
   path-ordered products $U(C_{xy})$ and $U(C_{(x+\mu)x}C_{xy})$.  The color
   indices are contracted according to the arrows. } \end{description}

\vspace{1.5cm}
\hspace{0.5cm}
\unitlength=1.00mm
\linethickness{0.5pt}
\begin{picture}(100.00,120.00)
\put(35.00,61.00){\oval(18.00,2.00)[r]}
\put(30.00,62.00){\line(0,1){14.00}}
\put(28.00,60.00){\line(0,1){18.00}}
\put(28.00,60.00){\line(1,0){7.00}}
\put(30.00,76.00){\line(1,0){14.00}}
\put(28.00,78.00){\line(1,0){14.00}}
\put(44.00,76.00){\line(0,1){20.00}}
\put(42.00,78.00){\line(0,1){18.00}}
\put(82.00,61.00){\oval(22.00,2.00)[l]}
\put(83.00,62.00){\oval(2.00,16.00)[t]}
\put(83.00,60.00){\oval(2.00,28.00)[b]}
\put(84.00,62.00){\line(1,0){14.00}}
\put(84.00,60.00){\line(1,0){16.00}}
\put(98.00,62.00){\line(0,1){50.00}}
\put(100.00,89.00){\oval(28.00,2.00)[r]}
\put(100.00,60.00){\line(0,1){28.00}}
\put(100.00,90.00){\line(0,1){22.00}}
\put(42.00,97.00){\oval(78.00,2.00)[l]}
\put(84.00,96.00){\line(0,1){16.00}}
\put(84.00,112.00){\line(1,0){14.00}}
\put(44.00,96.00){\line(1,0){40.00}}
\put(82.00,98.00){\line(-1,0){40.00}}
\put(100.00,113.00){\oval(68.00,2.00)[r]}
\put(82.00,98.00){\line(0,1){16.00}}
\put(82.00,114.00){\line(1,0){18.00}}
\put(133.00,118.00){\makebox(0,0)[cb]{$y$}}
\put(3.00,102.00){\makebox(0,0)[cb]{$x$}}
\put(77.00,103.00){\makebox(0,0)[rb]{$C_{xy}$}}
\put(63.00,91.00){\makebox(0,0)[ct]{$\Gamma_{yx}$}}
\put(30.00,62.00){\line(1,0){5.00}}
\end{picture}

\vspace{-4cm}

\begin{description} {\small
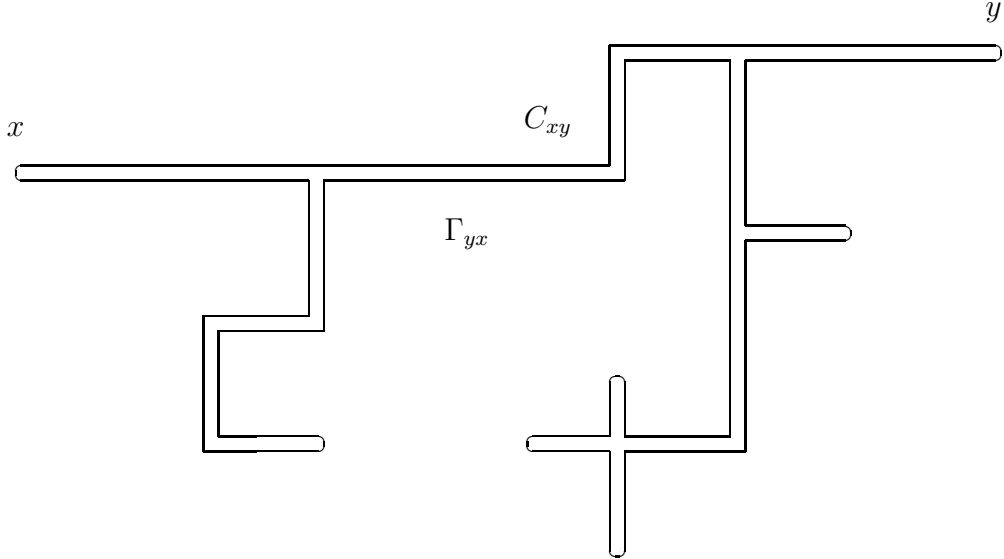
   \item[Fig. 5] \ \ \ The typical paths $\Gamma_{yx}$ which contribute the sum
   on the r.h.s.\ of \eq{path0} (and \eq{fpath0}). These $\Gamma_{yx}$ coincide
   with $C_{xy}$ passed backward modulo backtrackings which form a $1D$ tree.
   The sum over $\Gamma_{yx}$ is reduced to summing over the backtrackings.
   } \end{description}

\end{document}